\documentclass[prd,aps,showpacs,preprintnumbers,amsmath,nofootinbib,amssymb]{revtex4}
\pdfoutput=1 %added for JCAP compatibility
\usepackage{amsmath} 
\usepackage{amssymb} 
\usepackage{verbatim} 
\usepackage[pdftex]{graphicx}

\usepackage{dcolumn}% Align table columns on decimal point 
\usepackage{bm}% bold math 
\def\be{\begin{equation}} 
\def\ee{\end{equation}} 
\def\bea{\begin{eqnarray}} 
\def\eea{\end{eqnarray}} 
 
\begin{document} 
 
%\preprint{} 
 
%\date{\today} 
 
\title{The 21~cm Signature of Shock Heated and Diffuse Cosmic String Wakes} 
 
\author{Oscar F. Hern\'andez$^{1,2}$ 
and Robert H. Brandenberger$^{2}$
\email[email: ]{oscarh,rhb@physics.mcgill.ca}}
 
\affiliation{1) Marianopolis College, 4873 Westmount Ave., Westmount, QC H3Y 1X9, Canada \\
2) Department of Physics, McGill University, 
Montr\'eal, QC, H3A 2T8, Canada}

 
\begin{abstract} 

The analysis of the 21 cm signature of cosmic string wakes is extended 
in several ways. First we consider the constraints on $G\mu$ from the 
absorption signal of shock heated wakes laid down much later than 
matter radiation equality. Secondly we analyze the signal of diffuse wake, 
that is those wakes in which there is a baryon overdensity but which have 
not shock heated. Finally we compare the size of these signals to 
the expected thermal noise per pixel which dominates over the background 
cosmic gas brightness temperature and find that the cosmic string signal 
will exceed the thermal noise of an individual pixel in the 
Square Kilometre Array for string tensions $G\mu > 2.5 \times 10^{-8}$.

\end{abstract} 
 
\maketitle

\newcommand{\eq}[2]{\begin{equation}\label{#1}{#2}\end{equation}} 
 
\section{Introduction} 
\label{sec:introduction}
 
In a previous paper \cite{us} we studied the signal of a shock heated 
cosmic string wake in a position space 21cm redshift maps. 
We pointed out that cosmic string wakes give rise
to a pronounced signal in such maps, in particular at redshifts larger than
that of conventional reionization. The signal of a single wake is a wedge 
which is extended (a degree or larger for redshifts of $z\sim 20$) in both 
angular directions and thin in redshift direction. 
This effect comes from the fact that wakes form overdensities in baryons already
at very high redshifts, and that these regions then lead to the extra 21cm
emission or absorption.

In a followup paper \cite{us2} we computed the angular power spectrum for 
shock heated cosmic string wakes emitting 21 cm radiation. Because of our 
interest in shock heating in both \cite{us} and \cite{us2}
we focused our attention on those wakes with the greatest kinetic gas 
temperature $T_K$. Thus we concentrated our analyses on the wakes 
formed earliest, $z_i \sim 3000$ and emitting 21~cm radiation as late 
as possible before reionization, $z_e \sim 20$. This led us to miss the 
constraints on $G\mu$ that shock heated wakes formed at later 
$z_i$ could provide by considering the absorption of 21~cm radiation. 
In this paper we consider these constraints.

Furthermore, even if baryons collapsing onto the primordial string wake 
do not shock
heat, the string will nevertheless induce an overdense region of baryons.
This region, however, will be more diffuse than if shock heating occurs.
The wakes will be thicker but the overdensity smaller. In this paper, we 
study the 21cm signal of such a ``diffuse" cosmic string wake.

Current constraints on the cosmic string
tension from analyses of the angular power spectrum of CMB anisotropy
maps yield 
\be \label{limit1}
G \mu < 1.5 \times 10^{-7}
\ee
using combined data from WMAP and SPT microwave experiments \cite{Dvorkin}.
Limits based only on WMAP data gives a bound larger than this by a factor 
of two~\cite{previous}. It must be emphasized that any limits on $G \mu$ coming
from power spectra analyses implicitly depend on parameters
describing the cosmic string scaling solution which are quite
uncertain. Two prime examples of such parameters are the number
$N$ of string segments crossing any Hubble volume, and the
curvature radius $c_1$ of a long string relative to the Hubble radius.
These parameters are known only to within an order of magnitude
since they must be determined via numerical simulations of cosmic
string network evolution, and these simulations are extremely
challenging because of the huge hierarchy of scales which must
be included at the same time.

In our analyses below we show that a cosmic string signal will exceed 
the thermal noise of an individual pixel in a future radio telescope 
such as the Square Kilometre Array (SKA) for a string tension 
\be \label{limit2}
G \mu > 2.5\times 10^{-8} \, .
\ee
Our new limit is
independent of the parameters $N$ and $c_1$. 
In addition, since the
signal of a string wake has a very special geometry in position space,
position space shape detection algorithms are likely to be able to
detect string wake signals even if they do not stick out above the
Gaussian noise on a pixel by pixel basis. The situation is similar
to what is encountered when searching for cosmic string signals
in CMB anisotropy maps: since long string segments produce line
discontinuities in the maps, these can be identified by edge
detection algorithms such as the Canny algorithm \cite{Canny}
for values of the tension substantially lower than the value for which
the string signal dominates on a pixel by pixel basis, as has been
studied in detail in \cite{Amsel}.

The paper is organized as follows: 
We begin in Section \ref{sec:csreview} with a brief review of cosmic 
strings, their wakes, and their 21 cm signatures. Readers familiar 
with such results may skip this section. 
Section \ref{sec:diffuse-wakes} is the main part of our work. In this 
section we extend the analysis of the 21 cm signature of cosmic string 
wakes in several ways. First we consider the constraints 
on $G\mu$ from the absorptions signal of shock heated wakes laid down 
much later than matter radiation equality. Secondly we analyze the 
signal of diffuse wakes, that is those wakes in which there is a baryon 
overdensity but which have not shock heated. Finally we compare the 
size of these signals to the expected thermal noise per pixel 
which dominates over the  background cosmic gas brightness temperature. 
In Section~\ref{sec:conclusion} we present our conclusions and put our 
work in context. 

\section{Cosmic Strings, Their Wakes and 21 cm Signatures}
\label{sec:csreview}

Many particle physics models beyond the Standard Model predict the
existence of cosmic strings. In particular, cosmic strings are produced
at the end of inflation in many models of brane inflation \cite{Dvali} and
in many supergravity models \cite{Rachel}. Cosmic strings may also survive
in early universe models based on superstring theory such as ``string gas
cosmology" \cite{BV, RHBSGCrev}.  
Since the amplitude of cosmological signatures of cosmic strings is proportional
to the string tension, which itself is set by the energy scale of the new
physics, searching for signatures of strings in cosmology is a way of
probing Beyond the Standard Model physics which is complementary to
accelerator probes such as the Large Hadron Collider experiments which
probe the low energy limit of such theories.

Cosmic strings can be used as a cosmological
probe of New Physics because any particle physics model which admits 
stable cosmic strings as a solution inevitably results in a a network of 
such strings forming  during a symmetry breaking phase transition in the 
early universe (see \cite{ShellVil, HK, RHBrev} for reviews on cosmic 
strings and cosmology).
Such a network will persist to the present time
and approaches a dynamical attractor configuration, the 
so-called ``scaling solution" in which the statistical properties of
the string network are independent of time if all lengths are scaled to
the Hubble radius. The cosmic string scaling solution is charaterized by
a random-walk-like network of infinite (or ``long") strings with step length 
comparable to the Hubble radius, plus a distribution of cosmic string loops 
with radii smaller than the Hubble radius. The typical transverse
velocity of a long string segment is the speed of light. This implies
that at all times $t > t_{eq}$  ($t_{eq}$ being the time of equal matter and 
radiation) relevant for structure formation there are
strings which act as seeds for the accretion of matter.
The key parameter which characterizes cosmic strings and their cosmological
consequences is the mass per unit length $\mu$.  
It is usually quoted in terms of the dimensionless
number $G \mu$, where $G$ is Newton's gravitational constant. 

Both long string segments and string loops lead to nonlinearities in the
matter distribution at all redshifts up to $z_{eq}$,
the redshift of equal matter and radiation. Cosmic string network
simulations \cite{CSsimuls} indicate that the long strings have a larger 
effect than the loops, so we will here focus our attention on the effects of
long string segments, as we did in \cite{us, us2} (For a
recent study of the 21cm signal of a cosmic string loop see \cite{Pagano}.). 
In the standard $\Lambda$CDM cosmology without strings, nonlinear structures are
negligible until shortly before the time of reionization. Hence, the signals
of cosmic strings should be more and more pronounced against the ``noise"
of effects from structure formation in the $\Lambda$CDM model the
higher the redshift is.  21cm surveys hence appear as an ideal window to probe 
for the possible existence of cosmic strings. They probe the
distribution of matter at much higher redshifts than galaxy redshift
surveys since they map out the distribution of neutral
hydrogen in the universe, in particular before the time of reionization.
Compared to CMB maps, 21cm maps have the advantage of yielding
three-dimensional information.

The nonlinearity in the matter distribution induced by a long string
segment at time $t$ takes the form of a ``wake" \cite{CSwake} whose length 
is set by the length of the string segment (which will be similar to the 
Hubble radius $t$),
whose width is set by $v_s \gamma_s t$, where $v_s$ is the transverse velocity
of the string segment and $\gamma_s$ is the associated relativistic gamma
factor, and whose average thickness is proportional to $G \mu t$. The wake 
formation starts with an initial velocity perturbation and then grows by 
gravitational accretion. 

\begin{figure}[htbp]
\includegraphics[height=5cm]{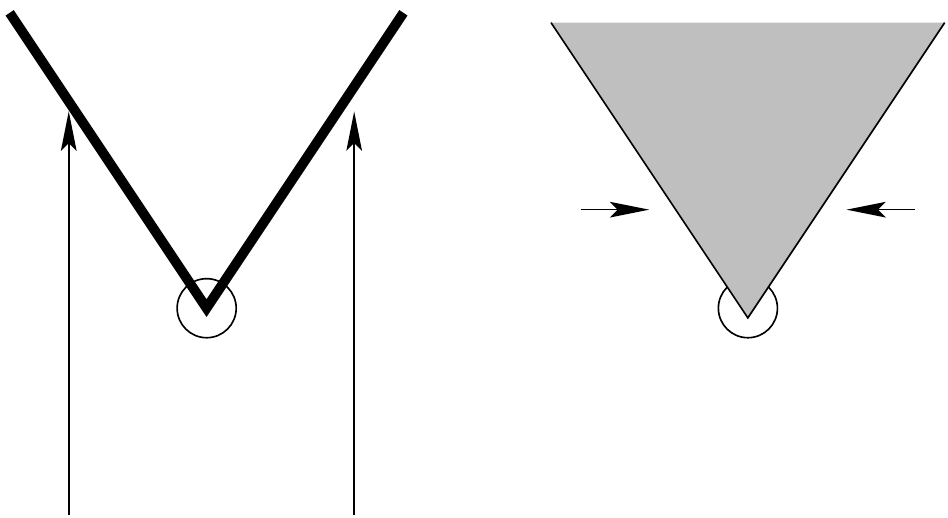}
\caption{Sketch of the mechanism by which a wake behind a moving
string is generated.  Consider a string perpendicular to the plane of
the graph moving straight downward. From the point of view of the
frame in which the string is at rest, matter is moving upwards, as
indicated with the arrows in the left panel. From the point of view
of an observer sitting behind the string (relative to the string motion)
matter flowing past the string receives a velocity kick towards the
plane determined by the direction of the string and the velocity
vector (right panel). This velocity kick towards the plane leads
to a wedge-shaped region behind the string with twice the
background density (the shaded region in the right panel).} 
\label{fig:wakewedge}
\end{figure}

The velocity perturbation (see Figure~\ref{fig:wakewedge}) stems from 
the fact that space perpendicular to a long string segment is conical, 
i.e. flat space with a missing
wedge whose deficit angle is $\alpha = 8 \pi G \mu$. 
A string moving through a gas of particles with transverse velocity $v_s$
will hence induce a velocity perturbation $\delta v$ of the gas towards 
the plane behind the string, whose magnitude is given by
$\delta v = 4 \pi G \mu v_s \gamma_s $.
.

A cosmic string segment laid down at time $t_i$ (we are interested
in $t_i \geq t_{eq}$) will generate a wake with
physical dimensions:
\be  \label{size}
 l_1(t_i)\times l_2(t_i)\times w(t_i)~ =~ t_i ~c_1 ~\times~ t_i ~v_s\gamma_s ~\times~ t_i ~4\pi G\mu  v_s\gamma_s \, .
\ee
where $c_1$ is a constant of order one. In the above, the first 
dimension is the length 
in direction of the string, the second is the depth, and the third is the
average width. After being laid down, the wake will grow by gravitational
accretion.  This has been studied using the Zel'dovich approximation \cite{Zel}
in a number of works \cite{wakegrowth}. In this method of analysis, we
follow the time evolution of the height above the centre of the wake of mass
shells to determine the comoving distance $q_{nl}$ which ``turns around'' 
at time $t$, i.e. for which the physical height $h(q, t)$ is maximal. We 
obtain that the comoving thickness grows linearly with the scale factor 
of the universe. A wake laid down at redshift $z_i$ will have comoving 
thickness at redshift $z$ given by
\be \label{qnl}
q_{nl}(z) \, = \, 
\frac{16 \pi}{5} {G \mu v_s \gamma_s\over H_o} (z_i  + 1)^{1/2}{1 \over (z+1)}
\, .
\ee
The physical height of the shell above the centre of the wake at the time of
``turnaround'' is half what the height would be in the absence of gravitational
accretion, and hence the relative overdensity is a factor of 2. It is then
assumed that the shell collapses to half of the height it had at turnaround,
that it then undergoes shocks and virializes at that distance. Then,
the overdensity will be a factor of 4 greater than the background density. 
This approximate analytical analysis has been confirmed in detailed
hydrodynamical simulations \cite{Sorn}. 

During the infall onto the string wake, the baryons acquire kinetic energy
which during the process of shock heating transforms to thermal energy.
The kinetic temperature $T_K$ acquired by the hydrogen atoms can be obtained 
from the kinetic energy which particles acquire when they hit the shock 
front. This is determined 
using the Zel'dovich approximation \cite{Zel}.The result obtained in 
\cite{us} is
\be
T_K \, \simeq \, [20~{\rm K}] (G \mu)_6^2 (v_s \gamma_s)^2 \frac{z_i + 1}{z + 1}  \, ,
\ee
where $(G \mu)_6$ indicates the value of $G \mu$ in units
of $10^{-6}$. The result is expressed in degrees Kelvin.

Note that the wake kinetic temperature increases as $z$ decreases since
the wakes have more time to grow, which in turn leads to a growing
gravitational potential. Since $T_{\gamma}$ decreases in time, then it
is more and more likely that the string wake 21cm signal will be in
emission rather than absorption as $t$ increases. However, it turns
out that for reshifts significantly larger than that of conventional
reionization and for string tensions smaller than the current limit
(\ref{limit1}), $T_K$ is smaller than $T_{\gamma}$ and that hence the string
wake signal will be in absorption.

The 21cm signal of a cosmic string wake has a distinctive geometry. 
We will see the string-induced absorption (or emission) regions in
directions of the sky where our past light cone intersects the
cosmic string wake. These are wedges in the sky of angular scale
determined by the comoving Hubble radius at the time that the string
is formed. The string wake grows in thickness, but its planar
dimensions are constant in comoving coordinates. For wakes formed
at a redshift corresponding to recombination this angular scale is
about one degree, for wakes formed at equal matter and radiation slightly
smaller. Since photons emitted at the top and bottom of the wake 
undergo slightly different redshifts, the region in 21cm redshift
surveys of extra emission/absorption takes on the form of a thin
wedge in redshift direction with a thickness which is \cite{us}
\be \label{freqwidth}
\left|{\delta z\over z+1}\right| =
\left|\frac{\delta \nu}{\nu}\right|
 \, = \, \frac{24 \pi}{15} G \mu v_s \gamma_s 
{\bigl( z_i  + 1 \bigl)^{1/2}\over \bigl( z+ 1 \bigr)^{1/2}} \, ,
\ee

\section{Diffuse Cosmic String Wakes}
\label{sec:diffuse-wakes}

The analysis in the previous section neglected the temperature of the gas
at the time that the wake was formed. The accretion onto a
string wake further increases the intrinsic gas velocities,
leading to an effective gas temperature which is $2.5 T_g$,
where $T_g$ is the temperature of the background gas. If 
\be
T_K \, < \, 2.5 \times T_g \, ,
\ee
then the incoherent velocities due to the thermal motion of
the accreted gas dominate over the coherent velocity induced
by accretion. In this case, there is no shock heating, and
the thickness of the overdense region induced by the wake
is larger than it would be in the absence of thermal motion
of the gas particles. The resulting overdense region we will
call a ``diffuse wake''.

We now estimate the size of the diffuse wake. In
linear cosmological perturbation theory, the total mass
accreted by the string wake is the same both in the
cases with and without shock heating. The difference
is that the large intrinsic thermal gas velocity will
render the wake wider and hence less dense. 
 
The first step is to compute the width $h_w$ of the wake
obtained taking into account the gas temperature $T_g$.
The result is
\be \label{height}
h_w(z)|_{T_K < T_g} \, = \, h_w(z)|_{T_g = 0} ~\Bigl( \frac{T_g}{T_K} \Bigr)
\ee
for $T_K < T_g$. The diffuse wake will thus be thicker by the
ratio of temperatures $T_g / T_K$, but the density of baryons
will be smaller by the same factor. Hence, the width of the
21cm signal induced by a diffuse wake will be thicker in redshift
direction by the above ratio of temperatures compared to what is
given in (\ref{freqwidth}). On the other hand, the 21cm brightness
temperature will be smaller since it is proportional to the baryon
density which is smaller than in the case of a shock-heated wake. 

The formula (\ref{height}) for the width of a diffuse wake 
can be derived by assuming equipartition of
energy between thermal energy and potential energy which for $T_K \ll T_g$ yields
\be \label{equip}
m_{HI} \delta \Phi \, = {3\over2}T_g \, ,
\ee
where $m_{HI}$ is the mass of a hydrogen atom and $\delta \Phi$ is
the gravitational potential induced by the wake. The latter, in
turn, is
\be \label{pot}
\delta \Phi \, = \, 2 \pi G \sigma |h| \, ,
\ee
where $\sigma$ is the surface density of the wake, and $|h|$ is
the height. Combining (\ref{equip}) with (\ref{pot}) yields
the linear scaling in temperature given in (\ref{height}).

The surface density $\sigma$ of a
diffuse wake is the same as that of a shock heated wake. It is given
by the energy per unit area of matter within a comoving height
$q_{nl}(z)$, the shell which is turning around at redshift $z$.
Thus $\sigma(z) = q_{nl}(z) \rho_0$,
%%
%\be
%\sigma(z) \, = \, q_{nl}(z) \rho_0 \, ,
%\ee
%%
where $\rho_o$ is the current background density. The
overdensity $\Delta \rho$ in the diffuse wake is given by
\be
\Delta \rho(z) \, = \, \frac{\sigma(z)}{h_w(z)} \, = \, \rho_0 ~\Bigl( \frac{T_K}{T_g} \Bigr)\, .
\ee
which results in
\be
{\rho\over\rho_0 } \, = \, \Bigl( 1+\frac{T_K}{T_g} \Bigr)\, .
\ee
The extra baryon density in the wake leads to extra 21~cm 
absorption or emission.  

The expression for the wake brightness temperature written in terms 
of kinetic temperature $T_{Kg}$ and the hydrogen atom number 
density $n_{HI}$ is very similar for a shock heated or a diffuse wake. The only difference is that $T_{Kg}$ is the kinetic temperature $T_K$ of the wake when shock heating occurs whereas it should be interpreted as the kinetic temperature of the cosmic gas $T_g$ for diffuse wakes.
\be \label{eq:deltaTb}
\delta T_b(z_e) \ = \, [17~{\rm mK}]\frac{x_c}{1+x_c}\left(1-\frac{T_\gamma}{T_{Kg}}\right)
\frac{n_{HI}^{wake}}{n_{HI}^{bg}} 
{(1+z_e)^{1/2}\over2\sin^2\theta}~.
\ee
Here, $z_e$ is the redshift of 21~cm emission or absorption, $T_\gamma$ is 
the CMB temperature, $\theta$ is the angle of the 21~cm ray with respect 
to the vertical to the wake, and $x_c$ are the collision coefficients whose 
values can be obtained from the tables listed in \cite{xc}. Here and 
throughout we take the cosmological parameters to be  
$H_0=73~{\rm km~s}^{-1}~{\rm Mpc}^{-1}$, 
$\Omega_b=0.0425$, $\Omega_m=0.26$. 
We work with a matter dominated universe for $z \le 3000$ with the age of 
the universe $t_0=4.3\times10^{17}$~s. The origin of the $\sin^{-2}(\theta)$
factor will be discussed in the Appendix, where it will also be shown that
this factor does not lead to any physical divergence.

For a shock heated wake, the ratio of the wake's hydrogen atom number density 
to the background density, $n_{HI}/n_{HI}^{bg}$, is 4. 
For a diffuse wake the relative brightness temperature is lower
than for a wake which has undergone shock heating since the number
density $n_{HI}$ of hydrogen atoms inside the diffuse wake is smaller,
and thus its ratio in (\ref{eq:deltaTb}) to the background density
$n_{HI}^{bg}$ is smaller. For our analysis below we take:
\bea
{n_{HI}\over n_{HI}^{bg}} 
 = \Big(1+{T_K\over T_g}\Big) ~~~~~~~~&~ & T_K\le 3 T_g \\ 
 =  4  ~~~~~~~~~~~~~~~~~~~~&~ & T_K \ge 3T_g
\eea
Note that for simplicity we have used $T_K \le 3 T_g$ instead of $T_K \le 2.5 T_g$.

The brightness temperature signal in (\ref{eq:deltaTb}) discussed in 
\cite{us,us2} was compared to the average brightness temperature of the 
surrounding cosmic gas. Between redshift $z=20$ and 35 the average background 
brightness temperature varies from -0.34~mK to -8.6~mK, respectively. 
However the analysis of the signal did not consider beam size nor thermal 
noise. In fact for $z\sim 20$ it is the thermal noise, and not the 
background brightness temperature, that  will first limit our ability to 
detect a cosmic string wake signal. The opposite is true for $z=35$. 
Beam size and thermal noise for the 21~cm brightness temperature
were considered in \cite{Oscar} where their Eq.~(4.19) gives the thermal 
noise per pixel:
\be
T_n=
\frac{12 ~\text{mK}~({\rm arcmin}/\theta_{\rm resolution})}{\sqrt{(\tau/10^4{\rm hr}) (A_e / {\rm km^2})}}
\Bigl(\frac{1+z}{21}\Bigr)^{3.85}
\Bigl[\frac{\sqrt{1+z}}{\sqrt{1+z}-1}\Bigr]^{1/2}
\label{Tnoiseperpixel}
\ee
Here $\tau$ is the total observing time and $A_e$ is the effective antenna area.
To derive this noise temperature per pixel in \cite{Oscar} we began with 
the thermal noise per visibility as given by Morales~\cite{Morales:2004ca} to arrive
at the thermal noise per pixel for the brightness temperature.
The angular resolution is $\lambda/D$ where $\lambda$ is the observation frequency and $D$ is the baseline. 
The angular resolution is assumed to be tuned by a dilution of the array
from being fully compact by a simple scaling of all baseline positions by
a fixed amount.  The system temperature is given by 
ARCADE 2~\cite{Fixsen:2009xn}. Further details can be found in ~\cite{Oscar}. 

The noise per pixel resolution dependence given in eq.~\ref{Tnoiseperpixel} 
assumes  an isotropic resolution in all three spatial directions. However 
for a cosmic wake signal, the width is order $4\pi G\mu(z_i+1)/(z+1)$ smaller 
than the two length directions, whereas each length direction is of order the 
Hubble size.  
If $\theta_{\rm hubble}(z)$ is the angular resolution needed for the wake's 
length, then $\theta_{\rm hubble}(z) 4\pi G\mu(z_i+1)/(z+1)$ is the angular 
resolution needed for the width. Thus an appropriate estimate of the 
resolution in radians needed to calculate the noise per pixel is:
\be
\theta_{\rm resolution} (z) =  \Big({4\pi G\mu(z_i+1)\over(z+1)}\Big)^{1/3} ~ \theta_{\rm hubble}(z)
= \Big({4\pi G\mu(z_i+1)\over(z+1)}\Big)^{1/3}  {1\over (\sqrt{z+1}-1)}
\ee

Hence, in a three dimensional map with the necessary resolution to detect a 
cosmic string wake laid down at $z_i$ and emitting or absorbing radiation at 
$z$, the thermal noise per pixel is:
\be
T_n^{\rm wake}=\frac{0.19~{\rm mK}~ (G\mu)_{9}^{-1/3}  (z_i+1)^{-1/3}  }
{\sqrt{(\tau/10^4{\rm hr}) (A_e / {\rm km^2})}}
\Big({z+1\over21}\Big)^{4.68}(1-1/\sqrt{z+1})^{1/2}
\ee
where $(G\mu)_{9}\equiv 10^{9} G\mu $. 

We will consider 10 000 hours of total observing time. Evidently for 
the SKA the effective antenna area is 1 sq km. For $z_i=3000$ and $z=20$, 
a $G\mu=4\times10^{-8}$ leads to an average noise per pixel 
$T_{n}=0.3~ {\rm mK}$ comparable in size to the background brightness 
temperature. Smaller $G\mu$ require a smaller resolution and lead to 
larger noise per pixel. Hence for $G\mu<4\times10^{-8}$ the thermal 
noise dominates over the background brightness temperature for the 
wake signal. For example at $z_i=3000$ and $z=20$
a $G\mu= 5\times10^{-9}$ gives a 
$\theta_{\rm resolution} (z)=5.8 \times 10^{-3}$ radians, which is 20 minutes 
of arc, and hence an average noise per pixel of $T_{n}=0.66~ {\rm mK}$. 
For strings laid down at a lower redshift the noise is greater.  
The noise is also a rising function of the emission redshift $z$. For a 
$z_i=1000$ with $z=20, 25$ and 35, $T_{n}=1.6~ {\rm mK},~2.0~ {\rm mK}$ 
and 2.7~ mK, respectively. 

\begin{figure}[htbp]
\includegraphics[height=6cm]{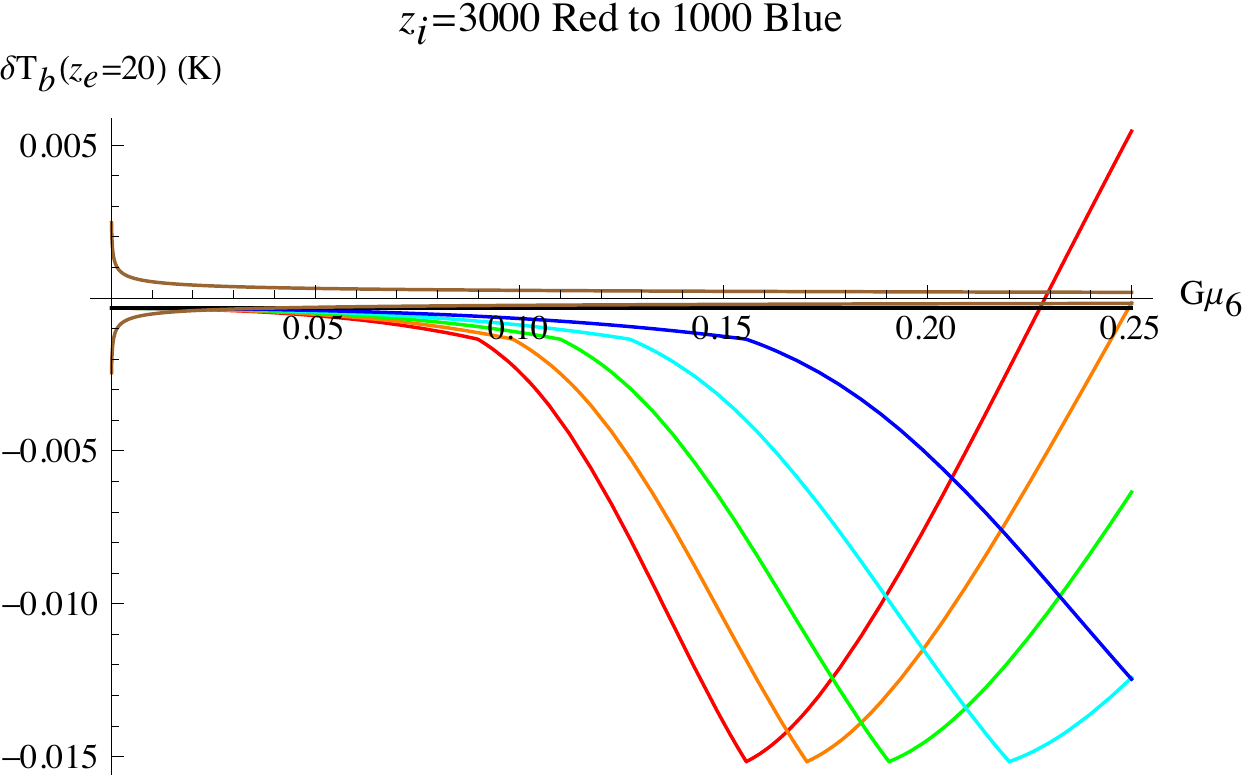}
\caption{The relative brightness temperature (vertical axis) in degrees 
Kelvin as a function of $(G \mu)_6$ (horizontal axis) for various values 
of the formation redshift $z_i$, evaluated for an observation redshift of 
$z = 20$. The curves from left to right (in the region of low values of 
$G \mu$) correspond to $z_i = 3000$ (red curve), $z_i = 2500$ (orange), 
$z_i = 2000$ (green), $z_i = 1500$ (light blue) and $z_i = 1000$ (dark blue). 
The two brown lines indicate the expected thermal noise per pixel in an 
experiment such as the SKA, with a pixel size chosen to depend on
$G \mu$ as described in the text. The black line that is almost 
indistinguishable from the x-axis is the brightness temperature of 
the background cosmic gas.} 
\label{fig:dTb20-vs-Gu}
\end{figure}

In Figure~\ref{fig:dTb20-vs-Gu} we show the relative brightness temperature 
(vertical axis) as a function of $(G \mu)_6$ (horizontal axis) for various
values of the redshift $z_i $ of wake formation, evaluated
at the redshift $z = 20$. Negative (positive) brightness temperature
means absorption (emission). The two almost horizontal (brown) lines 
give the noise per pixel level in an experiment such as the SKA for
the pixel size chosen such as to optimize the search for a string
wake with the respective value of $G \mu$ (see the above discussion).

\begin{figure}[htbp]
\includegraphics[height=6cm]{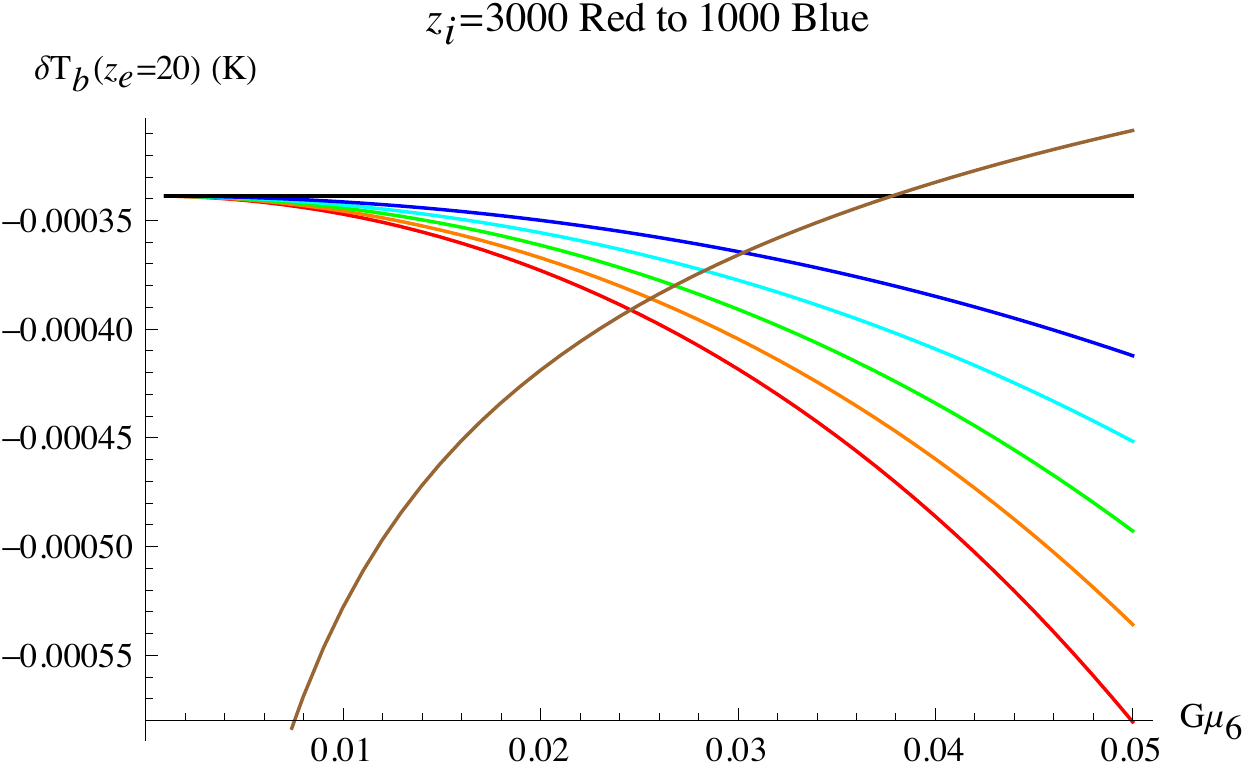}
\caption{A zoom in of figure~\ref{fig:dTb20-vs-Gu} to the point where 
the noise curve (brown) crosses the signal curves (colours red to blue). 
The black line at the top of the graph is the brightness temperature of 
the background cosmic gas.} \label{fig:dTb20-vs-Gu-Zoom}
\end{figure}

As expected, the larger the value of $z_i $, the easier it is
to detect low values of $G \mu$ since the wakes have had a longer
time to grow in thickness. A feature overlooked in our previous 
work~\cite{us, us2} is that for a fixed observational redshift $z$, there 
will be a range of values of $G \mu$ for which the wake's brightness 
temperature $T_K$ will be so close to the background cosmic gasp temperature 
that there will be no signal. In this situation,
we can consider smaller values of $z_i $ which lead to a clean and 
large absorption signal. 

For each curve in Figure~\ref{fig:dTb20-vs-Gu} it is apparent that 
there is a kink in
the brightness curve (in the absorption region). This kink occurs
at the value of $G \mu$ for which $T_K = 3 T_g$, the
transition point between a wake with shock heating (larger values
of $G \mu$) and a diffuse wake. Our previous analysis was only
valid until that point. We see here that with our current work
it is possible to extend the range of values of $G \mu$ which
can be probed by 21cm redshift surveys. 

Taking into account that the optimal pixel size will depend on
$G \mu$, then the pixel noise level will also depend on $G \mu$.
In Figure~\ref{fig:dTb20-vs-Gu-Zoom} we zoom in on the small
$G \mu$ region of the previous figure and compare the string
signal to the pixel noise. We see that the absorption
signal of the diffuse wakes is larger than
the noise level for values of $G \mu$ larger than
\be \label{limit3}
G \mu \,>\, 2.5 \times 10^{-8} \, 
\ee
for a detection redshift of $z = 20$. This is an
improvement of the current bound~(\ref{limit1})
and demonstrates the potential of 21cm surveys to probe
cosmic strings.

For smaller values of $G \mu$, cosmic string wakes will still
provide a signal. The signal has a particular shape in a three-dimensional
redshift survey and should hence be detectable using shape
detection algorithms for values of $G \mu$ much smaller than the
above value (\ref{limit3}). In work in progress we are exploring
this issue quantitatively. Note that a similar problem arises in
the case of identifying cosmic string signatures in CMB temperature
maps. It was shown \cite{Amsel} that edge detection algorithms provide
the promise of pushing the limits of detection several orders of
magnitude below the level of the pixel noise, and one order of
magnitude lower than limits which can be achieved using CMB angular
power spectra studies. Thus, we are optimistic that 21cm redshift
surveys will provide a very promising arena to probe for the
possible existence of cosmic strings.

\begin{figure}[htbp]
\includegraphics[height=6cm]{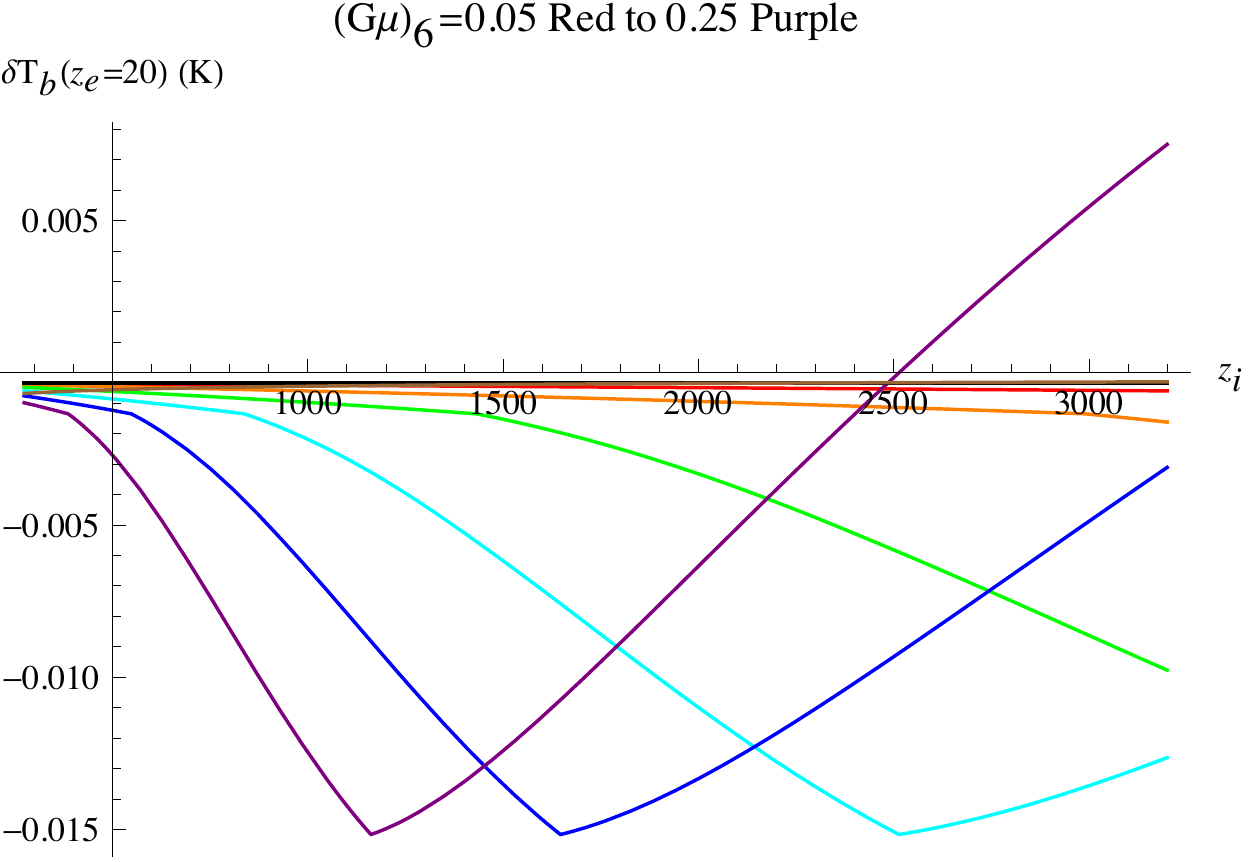}
\caption{The relative brightness temperature (vertical axis) as a 
function of formation redshift $z_i$ (horizontal axis) for various 
values of $(G \mu)_6$. The values of $(G \mu)_6$ chosen are 
0.05, 0.09, 0.13, 0.17, 0.21, and 0.25
(red, orange, green, light blue, dark blue and purple curves, respectively).
At low values of $z_i$ some of these curves are labelled by an increasing 
amplitude of the absorption signal. The black and brown lines near the 
x-axis represents the background gas brightness temperature and the noise 
temperature, respectively. They are nearly indistinguishable on this scale 
for these values of $G\mu$. The noise is plotted for $(G \mu)_6$ = 0.05. 
The larger values of $(G \mu)_6$ would give less noise.  } 
\label{fig:dTb20-vs-zi}
\end{figure}

In Figure~\ref{fig:dTb20-vs-zi} we show how that formation redshift 
which leads to the largest amplitude of the absorption signal depends 
on $z_i$. If we are looking
for strings of a particular value of $G \mu$ we would be focusing on wakes
formed at the redshift which yields largest absorption amplitude. Since
the formation redshift determines the size of the 21cm wedge which the
corresponding wake produces, we can use the size information to optimize
the search strategy. 

\section{Conclusions}
\label{sec:conclusion}

We have extended the analysis of the 21cm signal
of cosmic string wakes to the case when the string tension is too small to
yield shock heating of the wake. Instead, a ``diffuse wake" forms. This wake
is less dense but thicker than a wake which has undergone shock heating.
Hence, the relative 21cm brightness temperature is smaller in amplitude,
but it is extended over a larger redshift interval. 

Demanding that the relative brightness temperature exceeds the pixel
thermal noise in a future 21cm experiment such as the Square
Kilometre Array (SKA) (with the pixel size chosen to be able to
identify the wake signal, as described in the text) leads to a 
limit of $G \mu > 2.5 \times 10^{-8}$, an improvement over the
current limit of Eq.~(\ref{limit1}) obtained from CMB anisotropy 
maps~\cite{Dvorkin}. There are three important comments to make.
First, a limit obtained by direct searches in position space is 
independent of some of the unknown parameters characterizing the scaling
solution of strings (for example, it is largely independent of the number of
strings per Hubble volume), whereas a bound obtained from the angular
correlation function of CMB anisotropies depends on these parameters.
Thus, a bound obtained by the methods we describe is more robust.
Secondly with more statistics, even a signal strength below the noise 
can be detected.
And finally, since cosmic string wakes produce signals with a distinguished
geometry (namely thin wedges in redshift maps), these signals should
be detectable by clever statistical techniques even if the amplitude is
substantially smaller than the pixel noise. An example of a statistical
tool which could be used to look for the specific string wake signal is
the set of Minkowski functionals, statistics designed to characterize
the topology of the structure of maps. For a preliminary study of
this approach with references to the original literature see \cite{Evan}.

\section{Appendix}

In this appendix we present how the $(\sin{\theta})^{-2}$
factor in the expression for the wake brightness temperature arises from the line profile term $\phi(\nu)$.  We also explain why it does not imply any physical singularity in the limit $\theta \rightarrow 0$.

We begin with the formula for the relative brightness temperature of photons (see \cite{us} and references therein)
\be \label{three3}
\delta T_b(\nu) \, \simeq \, {\bigl( T_S - T_{\gamma} \bigr) \over (1+z)}
\tau_{\nu}  \, ,
\ee
%%
%%
%\be \label{three3}
%\delta T_b(\nu) \, \simeq \, T_S \frac{x_c}{1 + x_c} \bigl( 1 - \frac{T_{\gamma}}{T_K} \bigr) 
%\tau_{\nu} (1 + z)^{-1} \, ,
%\ee
%%
where the optical depth $\tau_{\nu}$ is given by
\be \label{three5}
\tau_{\nu} \, = \, 
\frac{3 c^2 A_{10}}{4 \nu^2} \bigl( \frac{\hbar \nu_{10}}{k_B T_S} \bigr) \frac{N_{HI}}{4} \phi(\nu) \, .
\ee
Here $N_{HI}$ is the column density of HI, and $\phi(\nu)$ is the line profile.

The line profile accounts for the nonzero width of the 21~cm line.
The line profile is a sharply peaked function about $\nu_{21}=$1420 MHz normalized such that
\be
{\int_0^\infty} \phi(\nu) d\nu =1
\ee
For the case at hand the line profile is not a Dirac delta function because the Hubble expansion leads to a velocity gradients along the line of sight. Due to this radial velocity gradient, the emitted 21~cm photons incur a local line broadening which one can interpret as a relative Doppler shift. 

\begin{figure}[htbp]
\includegraphics[height=8cm]{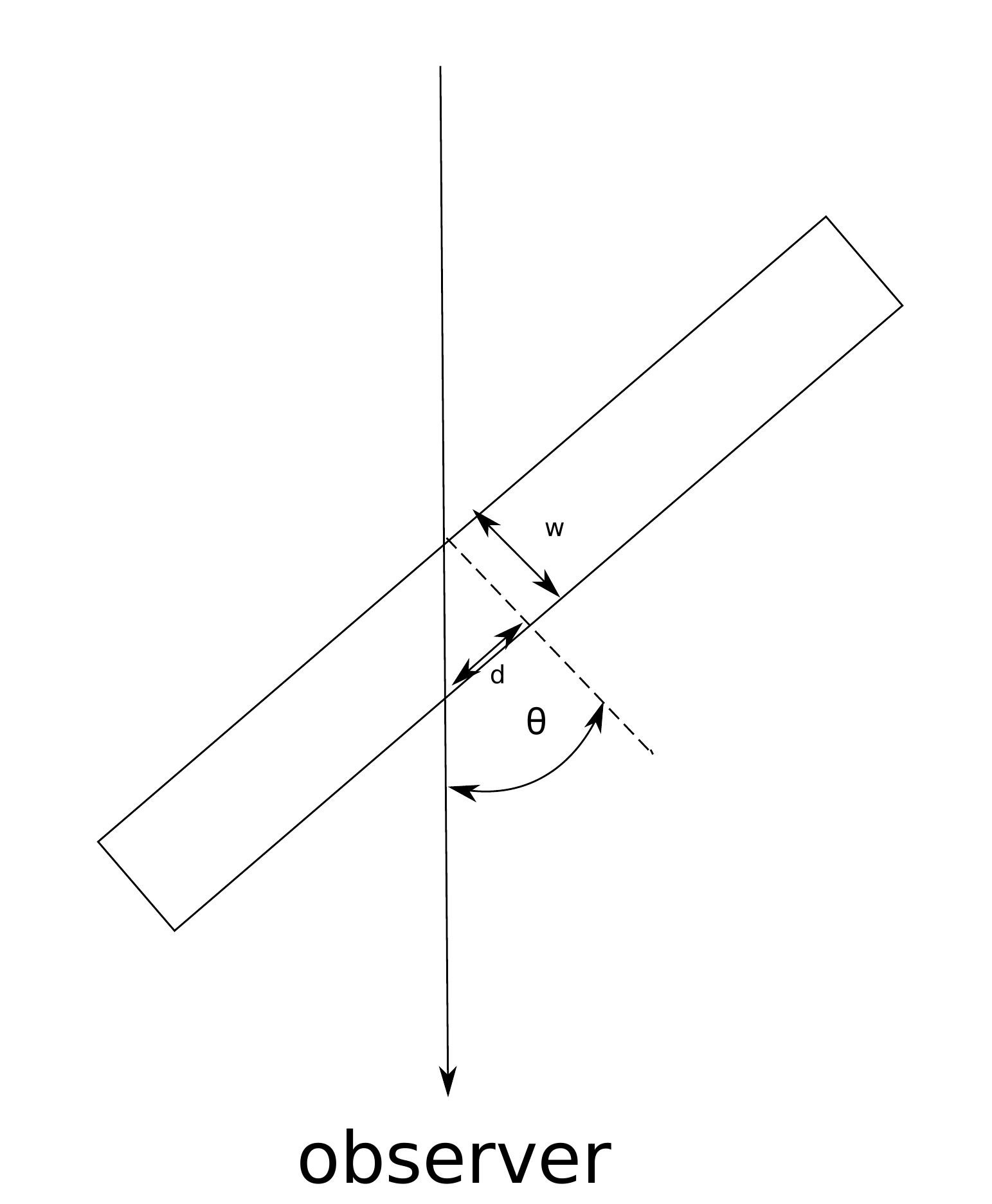}
\caption{A 21 cm light ray traverses a cosmic string wake of width $w$.} 
\label{fig:invsin2}
\end{figure}

In particular, let us consider photons reaching us at an angle $\theta$ relative
to the vertical of the wake having width $w$ (see fig.~\ref{fig:invsin2}). There is a relative Doppler shift $\delta \nu$
between photons from the front and back of the wake along the photon
trajectory. The horizontal distance $d$ along the wake between the back
point of the photon trajectory and the photon trajectory exiting the wake is
given by
\be
d \, = \, w \tan{\theta} \, .
\ee
The relative Doppler shift is due to the Hubble expansion of the planar
dimensions of the wake. Thus, the component of this expansion
velocity in direction of the photon trajectory is given by
\be
v_{//} \, = \, H d \sin{\theta} \, = \,{ H w \sin^2{\theta}\over \cos{\theta}} \, ,
\ee
and we have a frequency width given by
\be
\delta\nu=v_{//} ~ \nu_{21}
\ee
The line profile is
\be \label{three8}
\phi(\nu) \, = \, \frac{1}{\delta \nu} \,\,\, \rm{for} \,\,\, 
\nu \, \epsilon \, [\nu_{21} - \frac{\delta \nu}{2}, \nu_{21} + \frac{\delta \nu}{2}] \, ,
\ee
and $\phi(\nu) = 0$ otherwise. For a frequency within the
range of 21cm radiation passing through the wake the relative brightness
temperature is given by (\ref{eq:deltaTb}). This makes the origin
of the $(\sin{\theta})^{-2}$ factor manifest.

Any actual measurement will have a finite frequency resolution. If we integrate
over a frequency interval larger than $\delta \nu$, the $(\sin{\theta})^{-2}$ factor
will cancel as it must. If we have a very small frequency resolution $R$, then
the brightness temperature density at a fixed $\nu$ will increase as $\theta$
decreases until $\delta \nu$ becomes smaller than $R$. From that point
onwards, the result will be independent of $\theta$. Thus there is never any divergence associated with $\theta \rightarrow 0$.

\begin{acknowledgments} 
 
This work is supported in part by a NSERC Discovery Grant, by funds from the 
CRC Program, and by the FQRNT Programme de recherche 
pour les enseignants de coll\`ege. 

\end{acknowledgments}

\end{document}